\documentclass[reprint,amsmath,amssymb,onecolumn]{revtex4-1}

\usepackage{natbib}
\usepackage{graphicx}
\usepackage{dcolumn}
\usepackage{bm}

\makeatletter
\newcommand*{\rom}[1]{\expandafter\@slowromancap\romannumeral #1@}
\makeatother

\begin{document}

\title{Manifold spirals in barred galaxies with multiple pattern speeds}

 \author{C. Efthymiopoulos}
 \email{cefthim@academyofathens.gr}
\author{M. Harsoula}
 \email{mharsoul@academyofathens.gr}
\author{G. Contopoulos}%
 \email{gcontop@academyofathens.gr}
\affiliation{%
 Research Center for Astronomy and Applied Mathematics,
 Academy of Athens\\ Soranou Efesiou 4, GR-115 27 Athens, Greece}

\date{\today}

\begin{abstract}
In the manifold theory of spiral structure in barred galaxies, the usual assumption  
is that the spirals rotate with the same pattern speed as the bar. Here we generalize 
the manifold theory under the assumption that the spirals rotate with different 
pattern speed than the bar. More generally, we consider the case when one or more 
modes, represented by the potentials $V_2$, $V_3$, $\ldots$, co-exist in the galactic 
disc in addition to the bar's mode $V_{bar}$, but rotate with pattern speeds 
$\Omega_2$, $\Omega_3$, $\ldots$ incommensurable between themselves and with 
$\Omega_{bar}$. Through a perturbative treatment (assuming that $V_2,V_3...$ are 
small with respect to $V_{bar}$) we then show that the unstable Lagrangian points 
$L_1$, $L_2$ of the pure bar model $(V_{bar},\Omega_{bar})$ are `continued' in the 
full model as periodic orbits, when we have one extra pattern speed different from 
$\Omega_{bar}$, or as epicyclic `Lissajous-like' unstable orbits, when we have more 
than one extra pattern speeds. We denote by $GL_1$ and $GL_2$ the continued orbits 
around the points $L_1$, $L_2$, and we show that the orbits $GL_1$ and $GL_2$ are 
simply unstable. As a result, these orbits admit invariant manifolds which can be 
regarded as the generalization of the manifolds of the $L_1$,$L_2$ points 
in the single pattern speed case. As an example we compute the generalized orbits 
$GL_1$, $GL_2$ and their manifolds in a Milky-way like model with bar and spiral 
pattern speeds assumed different. We find that the manifolds produce a time-varying 
morphology consisting of segments of spirals or `pseudorings'. These structures are 
repeated after a period equal to half the relative period of the imposed spirals with 
respect to the bar. Along one period, the manifold-induced time-varying structures 
are found to continuously support at least some part of the imposed spirals, except 
at short intervals around those times at which the relative phase of the imposed 
spirals with respect to the bar becomes equal to $\pm\pi/2$. A connection of these 
effects to the phenomenon of recurrent spirals is discussed.
\end{abstract} 

\maketitle

\section{Introduction}
The manifold theory of spiral structure in barred galaxies (\cite{rometal2006}; 
\cite{vogetal2006}) predicts bi-symmetric spirals emanating from the end of galactic 
bars as a result of the outflow of matter connected with the unstable dynamics around 
the bar's Lagrangian points $L_1$ and $L_2$ (see also \cite{dan1965}). The manifold 
theory has been expressed in two versions, namely the `flux-tube' version 
(\cite{rometal2006}; \cite{rometal2007}; \cite{athetal2009a}; \cite{athetal2009b}; 
\cite{ath2012}) and the `apocentric manifold' version (\cite{vogetal2006}; 
\cite{tsouetal2008}; \cite{tsouetal2009}; \cite{haretal2016}). In both versions, 
the orbits of stars along the manifolds are chaotic, thus the manifolds provide 
a skeleton of orbits supporting `chaotic spirals' (\cite{pat2006}). Furthermore, 
the theory predicts that the orbital flow takes place in a direction preferentially 
{\it along} the spirals. This is in contrast to standard density wave theory 
(\cite{linshu1964}; see \cite{bintre2008}), which predicts a regular orbital flow 
forming `precessing ellipses' (\cite{kal1973}) which intersect the spirals. This 
difference has been proposed as an observational criterion to distinguish chaotic 
(manifold) from regular (density wave) spirals (\cite{pat2006}). 

Among a number of objections to manifold theory (see introduction in \cite{eftetal2019} 
for a review as well as \cite{fonetal2019}; \cite{diaetal2019} for more recent 
references), a common one stems from the long-recognized possibility that the bar and 
the spirals could rotate at different pattern speeds (\cite{selspa1988}; 
see \cite{bin2013}; \cite{sel2014} for reviews). Since the unstable equilibria 
$L_1$ and $L_2$ are possible to define only when the potential is static in a frame 
co-rotating with the bar, manifold spirals emanating from $L_1$ and $L_2$ are necessarily 
also static in the same frame, hence, they should co-rotate with the bar. This 
prediction seems hard to reconcile either with  observations (\cite{veretal2001}; 
\cite{booetal2005}; \cite{patetal2009}; \cite{meietal2009}; \cite{spewes2012}; 
\cite{antetal2014}; \cite{junetal2015}; \cite{speroo2016}) or simulations 
(\cite{selspa1988}; \cite{litcar1991}; \cite{rausal1999}; \cite{qui2003}; 
\cite{minqui2006}; \cite{dubetal2009}; \cite{quietal2011}; \cite{minetal2012}; 
\cite{babetal2013}; \cite{rocetal2013}; \cite{fonetal2014}; \cite{bab2015}). 
On the other hand, recently \cite{eftetal2019} found empirically that the manifold 
spirals, as computed in an N-body simulation by momentarily `freezing' the potential 
and making all calculations in a frame rotating with the instantaneous pattern speed 
of the bar, reproduce rather well the time-varying morphology of the N-body spirals; 
this, despite the fact that multiple patterns are demonstrably present in the latter 
simulation. Such a result points towards the question of whether manifold theory is 
possible to generalize under the presence of more than one patterns in the disc,  
rotating with different speeds. 

We hereafter present such a generalization of the manifold theory under the explicit 
assumption of multiple pattern speeds. In particular, we consider models of barred 
galaxies in which the disc potential at time $t$, considered in cylindrical 
co-ordinates $(\rho,\phi)$ in a frame corrotating with the bar, has the form
\begin{eqnarray}\label{potgen}
V_{disc}(\rho,\phi,t) &= &V_0(\rho)+V_{bar}(\rho,\phi)\nonumber\\ 
&+&V_2(\rho,\phi-(\Omega_2-\Omega_{bar})t) \\
&+&V_3(\rho,\phi-(\Omega_3-\Omega_{bar})t) +\ldots\nonumber
\end{eqnarray}
In such a  model, the bar rotates with pattern speed $\Omega_{bar}$, while $\Omega_2$, 
$\Omega_3$, etc. are the pattern speeds (incommensurable between themselves and with 
$\Omega_{bar}$) of additional non-axisymmetric perturbations, modelled by the potentials 
$V_2$, $V_3$, etc. The latter can be secondary spiral, ring, or bar-like modes, assumed 
to be of smaller amplitude than the principal bar. We can state our main result as 
follows: through Hamiltonian perturbation theory, we demonstrate that spiral-like 
invariant manifolds exist in the above generalized potential given by Eq.(\ref{potgen}). 
These manifolds emanate from special orbits which can be regarded as continuations of 
the unstable Lagrangian equilibria of the potential $V_0+V_{bar}$ after `turning on' 
the terms $V_2$, $V_3$ etc. Specifically, adding one more term $V_2$ with frequency 
$\Omega_2$, we can prove the existence of two {\it periodic} solutions in the bar's 
rotating frame, each of period equal to $\pi|\Omega_2-\Omega_{bar}|^{-1}$. These orbits, 
denoted hereafter as $GL_1$, $GL_2$, (standing for `generalized $L_1$ and $L_2$'), 
form epicycles of size $O(V_2)$ with a center near the bar's end, and they reduce to 
the usual Lagrangian points $L_1$ and $L_2$ when $V_2$ goes to zero. In the same way, 
adding two terms $V_2$ and $V_3$ with incommensurable frequencies $\Omega_2$, $\Omega_3$, 
allows to prove the existence of two {\it quasi-periodic} orbits (also denoted $GL_{1}$ 
and $GL_{2}$) reducing to two points $L_1$ and $L_2$ in the limit of $V_2$ ,$V_3$ both 
going to zero. Each of the orbits $GL_1$, $GL_2$ then appears as an epicyclic oscillation 
with the two frequencies $|\Omega_2-\Omega_{bar}|$, $|\Omega_3-\Omega_{bar}|$, thus 
forming a Lissajous figure around $L_1$ or $L_2$. One can continue in the same way 
adding more frequencies. The key result, shown in section 2 below, is that, independently 
of the number $M$ of assumed extra frequencies, the orbital phase space in the neighborhood 
of the generalized orbits $GL_1$ and $GL_2$ admits a decomposition into a {\it center
$^{M+1}\times$ saddle} linearized dynamics (see \cite{gometal2001}). Hence, the orbits 
$GL_1$ and $GL_2$ possess stable and unstable manifolds, which generalize the manifolds 
of the points $L_1$ and $L_2$ of the pure bar model. In particular, the unstable manifolds 
of the orbits $GL_1$, $GL_2$ support trailing spirals and ring-like structures. In fact, 
these manifolds have a similar morphology as the manifolds of the $L_1$ and $L_2$ points, 
but they are no longer static in the frame co-rotating with the bar. In physical terms, 
the manifolds of the $GL_1$ and $GL_2$ orbits adapt their form in time periodically or 
quasi-periodically to follow the additional patterns present in the disc. An explicit 
numerical example of this behavior is given in section 3, referring to a Milky-way 
like model in which bar and spirals rotate at different pattern speeds. In this 
example we explicitly compute the orbits $GL_1$ and $GL_2$ as well as the manifolds 
emanating from them. Remarkably, despite using only a coarse fitting approach, 
the manifolds provide a good fit to the model's imposed spirals. Since the latter have 
a relative rotation with respect to the bar, one has to test this fitting at different 
phases of the displacement of the spirals with respect to the bar's major axis. We find 
that the fitting is good at nearly all phases except close to $\pm\pi/2$. A possible 
connection of this effect with the phenomenon of recurrent spirals is discussed. 

The paper is structured as follows: section 2 gives the general theory, i.e., existence 
of the generalized unstable Lagrangian orbits $GL_1$ and $GL_2$ and their manifolds 
under the presence of multiple pattern speeds. Section 3 presents our numerical example, 
in which the manifolds are constructed under a different pattern speed of the bar and 
the spirals. Section 4 gives the summary of results and conclusions. Mathematical 
details on the series computations described in section 2, using the Lie method, 
are given in the Appendix. 

\section{Theory}
The Hamiltonian in the disc plane in a galactic model with the potential 
(\ref{potgen}) can be written as
\begin{equation}\label{hamgen}
H = H_0+H_1 
\end{equation}
where $H_0$ is the axisymmetric + bar Hamiltonian:
\begin{equation}\label{hambar}
H_0={1\over 2}\left(p_\rho^2 + {p_\phi^2\over 2\rho^2}\right) 
- \Omega_{bar}p_\phi + V_0(\rho) + V_{bar}(\rho,\phi)
\end{equation}
and $H_1 = V_2+V_3+...$. The pair $(\rho,\phi)$ are the test particle's 
cylindrical co-ordinates in a frame rotating with angular speed $\Omega_{bar}$, 
while $(p_\rho,p_\phi)$ are the values of the radial velocity and angular momentum 
per unit mass of the particle in the rest frame. The dependence of the Hamiltonian 
$H$ on time can be formally removed by introducing extra action-angle pairs. 
Setting the angles $\phi_2=(\Omega_2-\Omega_{bar})t$, $\phi_3=(\Omega_3-\Omega_{bar})t$, 
etc., with conjugate dummy actions $I_2$, $I_3$, etc., we arrive at the extended 
Hamiltonian 
\begin{eqnarray}\label{hamext}
H &=&  {1\over 2}\left(p_\rho^2 + {p_\phi^2\over 2\rho^2}\right) 
- \Omega_{bar}p_\phi +V_0(\rho) + V_{bar}(\rho,\phi)\nonumber\\
&+& (\Omega_2-\Omega_{bar})I_2 +(\Omega_3-\Omega_{bar})I_3 +...\\
&+&V_2(\rho,\phi-\phi_2)+V_3(\rho,\phi-\phi_3)+\ldots\nonumber
\end{eqnarray}
which yields the same equations of motion as the Hamiltonian (\ref{hamgen}). 

The Hamiltonian $H_0$ gives rise to the two Lagrangian equilibrium points: 
$L_1 = (\rho_{L1},\phi_{L1},p_{\rho,L1}=0,p_{\phi,L1}=\Omega_{bar}\rho_{L1}^2)$ and 
$L_2 = (\rho_{L2},\phi_{L2},p_{\rho,L2}=0,p_{\phi,L2}=\Omega_{bar}\rho_{L2}^2)$ such 
that 
$\partial H_0/\partial\rho=
\partial H_0/\partial\phi=
\partial H_0/\partial p_\rho=
\partial H_0/\partial p_\phi=0$ at the points $L_1$ and $L_2$. Focusing on, say, $L_1$, 
and defining 
$\delta\rho = \rho-\rho_{L1}$, $\delta\phi=\phi-\phi_{L1}$, $J_\phi=p_\phi-p_{\phi,L1}$, 
the Hamiltonian $H_0$ can be expanded around the phase-space co-ordinates of the point 
$L_1$. This yields $H_0 = const + H_{0,2}+ H_{0,3}+...$, where $H_{0,2}$, $H_{0,3}$, 
$\ldots$ are quadratic, cubic, etc. in the variables 
$(\delta\rho,\delta\phi,p_\rho,J_\phi)$. Then, by a standard procedure 
(see the Appendix) we can define a linear transformation 
\begin{equation}\label{lintra}
\left(
\begin{array}{c}
\delta\rho\\
\delta\phi\\
p_\rho\\
J_\phi
\end{array}\right)
=
{\cal A}\cdot
\left(\begin{array}{c}
u\\
Q\\
v\\
P\\
\end{array}\right)
\end{equation}
where ${\cal A}$ is a $4\times 4$ matrix with constant entries, such that in the 
new variables $(u,Q,v,P)$ the quadratic part of the Hamiltonian $H_0$ takes 
a diagonal form
\begin{equation}\label{h02new}
H_{0,2} = \nu uv + {\kappa\over 2} (Q^2+P^2) 
\end{equation}
with $\nu,\kappa$ real constants. The matrix ${\cal A}$ satisfies the symplectic 
condition ${\cal A} \cdot {\cal J} \cdot {\cal A}^T = {\cal A}^T \cdot {\cal J} 
\cdot {\cal A} = {\cal J}$, where ${\cal J}$ is the $4\times 4$ fundamental 
symplectic matrix. The constants $\nu$, $\kappa$ are related to the eigenvalues of 
the variational matrix
\begin{equation}\label{varmat}
{\cal M} = 
\left(\begin{array}{cccc}
{\partial^2 H_0\over\partial\rho^2}~~
{\partial^2 H_0\over\partial\rho\partial\phi}~~
{\partial^2 H_0\over\partial\rho\partial p_\rho}~~
{\partial^2 H_0\over\partial\rho\partial p_\phi} \\
{\partial^2 H_0\over\partial\phi\partial\rho}~~
{\partial^2 H_0\over\partial\phi^2}~~
{\partial^2 H_0\over\partial\phi\partial p_\rho}~~
{\partial^2 H_0\over\partial\phi\partial p_\phi} \\
{\partial^2 H_0\over\partial p_\rho\partial\rho}~~
{\partial^2 H_0\over\partial p_\rho\partial\phi}~~
{\partial^2 H_0\over\partial p_\rho^2}~~
{\partial^2 H_0\over\partial p_\rho\partial p_\phi} \\
{\partial^2 H_0\over\partial p_\phi\partial\rho}~~
{\partial^2 H_0\over\partial p_\phi\partial\phi}~~
{\partial^2 H_0\over\partial p_\phi\partial p_\rho}~~
{\partial^2 H_0\over\partial p_\phi^2}
\end{array}\right)
\end{equation}
evaluated at the point $L_1$ via the relations $\lambda_{1,3} = 
\pm\nu$, $\lambda_{2,4} =\pm i\kappa$. Furthermore. the columns of the matrix 
${\cal A}$ are derived from the unitary eigenvectors of ${\cal M}$ (see the Appendix). 
Finally, the constant $\kappa$ is equal to the epicyclic frequency at the distance 
$\rho_{L1}$, namely $\kappa^2 = \partial^2 V_0/\partial\rho_{L1}^2 + 
3p_{\phi.L1}^2/\rho_{L1}^4$ (assuming $V_0(\rho)$ in Eq.(\ref{potgen}) to represent  
the entire disc's axisymmetric potential term, i.e., $<V_{bar}>=0$, where the average 
is taken with respect to all angles at fixed $\rho$). 

The Hamiltonian $H_{0,2}$ in Eq.(\ref{h02new}) describes the linearized dynamics 
around $L_1$: the harmonic oscillator part $(\kappa/2)(Q^2+P^2)$ describes epicyclic 
oscillations with the frequency $\kappa$, while the hyperbolic part $\nu u v$ implies 
an exponential dependence of the variables $u$ and $v$ on time, namely 
$u = u_0 e^{\nu t}$, $v = v_0e^{-\nu t}$. The linearized phase space can be decomposed 
in three subspaces, namely the invariant plane $E^C_{L1}=(Q,P)$, called the `linear 
center manifold', as well as the axes $u$, called the linear unstable manifold 
$E^U_{L1}$, and $v$, called the linear stable manifold $E^S_{L1}$ of the point $L_1$. 
The linearized equations of motion yield independent motions in each of the spaces 
$E^C_{L1}$, $E^U_{L1}$ and $E^S_{L1}$. Those on $E^U_{L1}$ describe orbits receding 
exponentially fast from $L_1$. A simple analysis shows that the outflow defined by 
such orbits has the form of trailing spiral arms. Basic theorems on invariant 
manifolds (\cite{gro1959}; \cite{har1960}) predict that the invariant subspaces 
$E^C_{L1}$, $E^U_{L1}$ and $E^S_{L1}$ of the linearized model are continued as 
invariant sets $W^C_{L1}$, $W^U_{L1}$, and $W^S_{L1}$, respectively, in the full 
nonlinear model given by the Hamiltonian $H_0$. In particular, the linear unstable 
manifold $E^U_{L1}$ is tangent, at the origin, to the unstable manifold $W^U_{L1}$ 
of the full model. The latter is defined as the set of all initial conditions tending 
asymptotically to $L_1$ when integrated backwards in time. In the forward sense of 
time, these orbits form an outflow which deviates exponentially from $L_1$. This 
outflow forms trailing spiral arms or ring-like structures which can be visualized 
either as `flux tubes' (\cite{rometal2006}), or as `apocentric manifolds' 
\cite{vogetal2006}. For more details and precise definitions see \cite{eft2010},  
\cite{eftetal2019} and references therein. 

We now extend the previous notions from the Hamiltonian $H_0$ to the full model 
of Eq.(\ref{hamext}). To this end, we first consider the canonical transformation 
(\ref{lintra}), which is defined only through the coefficients of the second order 
expansion around the co-ordinates of $L_1$ of the $H_0$ part of the Hamiltonian 
(i.e. the first line in Eq.(\ref{hamext})). Substituting this transformation to the 
full Hamiltonian, and omitting constants, the Hamiltonian takes the form:
\begin{eqnarray}\label{hamextuv}
H &= &\nu uv + {\kappa\over 2}(Q^2+P^2) \nonumber\\
&+& \sum_{s=3}^{\infty} 
\sum_{\substack{k_1,l_1,k_2,l_2\geq 0\\
k_1+k_2+l_1+l_2=s}} 
h_{k_1,k_2,l_1,l_2} u^{k_1} Q^{k_2} v^{l_1} P^{l_2}\nonumber\\
&+&(\Omega_2-\Omega_{bar}) I_2+(\Omega_3-\Omega_{bar}) I_3+...\\
&+& \sum_{s=1}^{\infty} 
\sum_{\substack{k_1,l_1,k_2,l_2\geq 0\\
k_1+k_2+l_1+l_2=s}}
V_{(2),k_1,k_2,l_1,l_2}(\phi_2) u^{k_1} Q^{k_2} v^{l_1} P^{l_2}\nonumber\\
&+& \sum_{s=1}^{\infty} 
\sum_{\substack{k_1,l_1,k_2,l_2\geq 0\\
k_1+k_2+l_1+l_2=s}}
V_{(3),k_1,k_2,l_1,l_2}(\phi_3) u^{k_1} Q^{k_2} v^{l_1} P^{l_2}+\ldots\nonumber
\end{eqnarray}
The functions $V_{(2),k_1,k_2,l_1,l_2}(\phi_2)$, $V_{(3),k_1,k_2,l_1,l_2}(\phi_3)$, etc 
are trigonometric in the angles $\phi_2$, $\phi_3$, etc., while $h_{k_1,k_2,l_1,l_2}$ 
are constants. The Lagrangian point $L_1$ has co-ordinates $u=v=Q=P=0$. This 
is no longer an equilibrium solution of the full Hamiltonian: by Hamilton's 
equations, one obtains, in general, $\dot{u}\neq 0$, $\dot{v}\neq 0$, 
$\dot{Q}\neq 0$, $\dot{P}\neq 0$ for $u=v=Q=P=0$, provided that at least one 
of the functions $V_{(2),k_1,k_2,l_1,l_2}(\phi_2)$, $V_{(3),k_1,k_2,l_1,l_2}(\phi_3)$, 
etc., are different from zero for $k_1+k_2+l_1+l_2=1$. However, the existence of an 
equilibrium solution of the Hamiltonian (\ref{hamextuv}) can be proven using 
perturbation theory. In particular, as a consequence of a theorem proven in 
\cite{gio2001}, there is a near-to-identity canonical transformation 
\begin{eqnarray}\label{uxitra}
&~&(u,Q,v,P,\phi_2,\phi_3,...,I_2,I_3,...)~~~~~~~~~~~~~~\\
&~&~~~~~~~~\rightarrow(\xi,q,\eta,p,\theta_2=\phi_2,\theta_3=\phi_3,...,J_2,J_3,...)
\nonumber
\end{eqnarray}
with 
\begin{eqnarray}\label{uxitra2}
u &=&\xi + F_u(\xi,q,\eta,p;\phi_2,\phi_3,\ldots) \nonumber\\  
Q &=&q + F_Q(\xi,q,\eta,p;\phi_2,\phi_3,\ldots) \\  
v &=&\eta + F_v(\xi,q,\eta,p;\phi_2,\phi_3,\ldots) \nonumber\\  
P &=&p + F_P(\xi,q,\eta,p;\phi_2,\phi_3,\ldots) \nonumber  
\end{eqnarray}
where the functions $F_u$, $F_Q$, $F_v$, $F_P$ are polynomial series of second 
or higher degree in the variables $(\xi,q,\eta,p)$ and trigonometric in the 
angles $\phi_2,\phi_3$ etc, such that, in the variables $(\xi,q,\eta,p,\theta_2,
\theta_3,...,J_2,J_3,...)$ the Hamiltonian (\ref{hamextuv}) takes the form 
\begin{eqnarray}\label{hamextxieta}
H &= &\nu \xi\eta + {\kappa\over 2}(q^2+p^2) \nonumber\\
&+& \sum_{s=3}^{\infty} 
\sum_{\substack{k_1,l_1,k_2,l_2\geq 0\\
k_1+k_2+l_1+l_2=s}} 
g_{k_1,k_2,l_1,l_2} \xi^{k_1} q^{k_2} \eta^{l_1} p^{l_2}\nonumber\\
&+&(\Omega_2-\Omega_{bar}) J_2+(\Omega_2-\Omega_{bar}) J_3+...\\
&+& \sum_{s=0,2,3,\ldots} 
\sum_{\substack{k_1,l_1,k_2,l_2\geq 0\\
k_1+k_2+l_1+l_2=s}} 
\Phi_{(2),k_1,k_2,l_1,l_2}(\phi_2) \xi^{k_1} q^{k_2} \eta^{l_1} p^{l_2}\nonumber\\
&+& \sum_{s=0,2,3,\ldots} 
\sum_{\substack{k_1,l_1,k_2,l_2\geq 0\\
k_1+k_2+l_1+l_2=s}} 
\Phi_{(3),k_1,k_2,l_1,l_2}(\phi_3) \xi^{k_1} q^{k_2} 
\eta^{l_1} p^{l_2}+\ldots\nonumber
\end{eqnarray}
The functions $\Phi_{(2),k_1,k_2,l_1,l_2}(\phi_2)$, 
$\Phi_{(3),k_1,k_2,l_1,l_2}(\phi_3)$, etc, are again trigonometric in the angles 
$\phi_2$, $\phi_3$, etc., while $g_{k_1,k_2,l_1,l_2}$ are constants. 

The formal difference between the Hamiltonians (\ref{hamextuv}) and 
(\ref{hamextxieta}) is the lack, in the latter case, of polynomial terms linear 
in the variables $(\xi,q,\eta,p)$. As a consequence, the point $\xi=q=\eta=p=0$ 
is an equilbrium point of the system as transformed to the new variables, 
since from Hamilton's equations for the Hamiltonian (\ref{hamextxieta}) 
one has $\dot{\xi}=\dot{q}=\dot{\eta}=\dot{p}=0$ if $\xi=q=\eta=p=0$. 
Using the transformation (\ref{uxitra2}) the equilibrium solution can be 
represented in the original variables as a `generalized $L_1$ solution':
\begin{eqnarray}\label{uxitral1}
u_{GL1} &=&F_u(0,0,0,0;\phi_2,\phi_3,\ldots) \nonumber\\  
Q_{GL1} &=&F_Q(0,0,0,0;\phi_2,\phi_3,\ldots) \\  
v_{GL1} &=&F_v(0,0,0,0;\phi_2,\phi_3,\ldots) \nonumber\\  
P_{GL1} &=&F_P(0,0,0.0;\phi_2,\phi_3,\ldots) \nonumber  
\end{eqnarray}
Through the linear transformation (\ref{lintra}) we obtain also the functions 
$\delta\rho_{GL1}(\phi_2,\phi_3...)$, $\delta\phi_{GL1}(\phi_2,\phi_3...)$, 
$p_{\rho,GL1}(\phi_2,\phi_3...)$, $J_{\phi,GL1}(\phi_2,\phi_3...)$. Since 
$\phi_2=(\Omega_2-\Omega_{bar})t, \phi_3=(\Omega_3-\Omega_{bar})t$, etc, the above 
functions determine the time-dependence of the original phase-space co-ordinates 
of the generalized trajectory $GL_1$. The trajectory depends trigonometrically on 
the phases $\phi_2,\phi_3,...$, hence it depends on time through the frequencies 
$|\Omega_2-\Omega_{bar}|$, $|\Omega_3-\Omega_{bar}|$ etc. In particular, the 
trajectory $GL_1$ is a {\it periodic orbit} (`a 1-torus') when there is one extra 
pattern speed. This generalizes to a Lissajous-like figure (M-torus) when there are 
$M>1$ extra pattern speeds etc. Through the linear part of Hamilton's equations for 
the Hamiltonian (\ref{hamextxieta}), we find that the equilibrium point $(\xi,q,\eta,p)$ 
is simply unstable (the variational matrix has one pair of real eigenvalues equal to 
$\pm\nu$ and one pair of imaginary eigenvalues equal to $\pm i\kappa$). Taking into 
account also the frequencies $|\Omega_2-\Omega_{bar}|$, $|\Omega_3-\Omega_{bar}|$, etc., 
the complete phase space in the neighborhood of the solution $GL_1$ can be decomposed 
into a $center^{M+1} \times saddle$ topology (\cite{gometal2001}). In particular:

-The phase-space invariant subset ${\cal W}^C_{GL1}$ defined by the condition 
$\xi=\eta=0$ is invariant under the flow of the Hamiltonian (\ref{hamextxieta}). 
It is hereafter called the `center manifold' of the orbit $GL_1$. Its dimension is 
$2+2M$, where $M$ is the number of additional frequencies. By the structure of 
Hamilton's equations, ${\cal W}^C_{GL1}$ is a {\it normally hyperbolic invariant 
manifold} (NHIM; see \cite{wig1994}).

-The set ${\cal W}^U_{GL1}$ of all initial conditions tending asymptotically to 
the generalized orbit $GL_1$ in the backward sense of time is the unstable manifold 
of the orbit $GL_1$. Basic theorems of dynamics (\cite{gro1959}; \cite{har1960}) 
guarantee that such an invariant manifold exists, and it is tangent, at the origin, 
to the linear unstable manifold ${\cal E}^U_{GL1}$, which coincides with the axis 
$\xi$ with $q=\eta=p=0$. Both ${\cal E}^U_{GL1}$ and ${\cal W}^U_{GL1}$ are 
one-dimensional. The product of ${\cal W}^U_{GL1}$ with the angles 
$\phi_2=(\Omega_2-\Omega_{bar})t$, $\phi_3=(\Omega_3-\Omega_{bar}) t$ etc, defines 
the {\it generalized unstable tube manifold} of the orbit $GL_1$, denoted hereafter 
as ${\cal W}^{TU}_{GL1}$. 

- Similar definitions hold for the stable manifold ${\cal W}^S_{GL1}$ and stable 
tube manifold ${\cal W}^{TS}_{GL1}$ of the orbit $GL_1$, which represent orbits 
tending asymptotically to the orbit $GL_1$ in the forward sense of time. 

As in the standard manifold theory of spirals, the basic objects giving rise to 
spirals are the generalized unstable tube manifolds ${\cal W}^{TU}_{GL1}$ and 
${\cal W}^{TU}_{GL2}$ of the orbits $GL_1$ and $GL_2$ respectively. A basic argument 
allows to show the following: {\it the projections of ${\cal W}^{TU}_{GL1}$ and 
${\cal W}^{TU}_{GL2}$ on the configuration space are trailing spirals emanating from 
the neighborhood of the bar's Lagrangian points $L_1$, $L_2$, but with a position 
and shape varying in time quasi-periodically. The variation is small and characterized 
by as many frequencies as the additional pattern speeds.} The argument 
is as follows: instead of the transformation (\ref{uxitra2}) one can formally 
compute a standard Birkhoff transformation (see \cite{eft2012}) of the form
\begin{eqnarray}\label{uxitrabir}
u &=&\xi_B + B_u(\xi_B,q_B,\eta_B,p_B;\phi_2,\phi_3,\ldots) \nonumber\\  
Q &=&q_B + B_Q(\xi_B,q_B,\eta_B,p_B;\phi_2,\phi_3,\ldots) \\  
v &=&\eta_B + B_v(\xi_B,q_B,\eta_B,p_B;\phi_2,\phi_3,\ldots) \nonumber\\  
P &=&p_B + B_P(\xi_B,q_B,\eta_B,p_B;\phi_2,\phi_3,\ldots) \nonumber  
\end{eqnarray}
such that the Hamiltonian (\ref{hamextuv}) expressed in the new variables 
$(\xi_B,q_B,\eta_B,p_B)$ becomes independent of the angles $\phi_2$, $\phi_3$, etc., 
namely, it takes the form (apart from a constant)
\begin{eqnarray}\label{hamextbir}
H &= &\nu \xi\eta_B + {\kappa\over 2}(q_B^2+p_B^2) \nonumber\\
&+& \sum_{s=1}^{\infty} 
\sum_{\substack{k_1,l_1,k_2,l_2\geq 0\\
k_1+k_2+l_1+l_2=s}} 
f_{k_1,k_2,l_1,l_2} \xi_B^{k_1} q_B^{k_2} \eta_B^{l_1} p_B^{l_2}~~.
\end{eqnarray}
Contrary to the normalization leading to the Hamiltonian (\ref{hamextxieta}), 
the Birkhoff normalization leading to the Hamiltonian (\ref{hamextbir}) is not 
guaranteed to converge (see \cite{eft2012}), thus it cannot be used to theoretically 
demonstrate the existence of the manifolds ${\cal W}^{TU}_{GL1}$, ${\cal W}^{TU}_{GL2}$. 
For practical purposes, however, the Birkhoff normalization can proceed up to 
an exponentially small remainder, hence the Hamiltonian (\ref{hamextbir}) 
approximates the dynamics with an exponentially small error. In this approximation, 
the coefficients $f_{k_1,k_2,l_1,l_2}$ are of the order of the amplitude of the 
extra patterns if $k_1+k_2+l_1+l_2\leq 2$, while one has 
$f_{k_1,k_2,l_1,l_2} = h_{k_1,k_2,l_1,l_2} + h.o.t$ if $k_1+k_2+l_1+l_2>2$. 
Hence, the resulting Hamiltonian is dominated by the bar terms. The equilibrium 
solutions representing the generalized Lagrangian equilibria $GL_1$ and $GL_2$ can 
be computed as the (non-zero) roots of Hamilton's equations $\dot{\xi}_B = 
\dot{q}_B=\dot{\eta}_B=\dot{p}_B=0$. The key remark is that, since the Hamiltonian 
(\ref{hamextbir}) no longer depends on time, {\it the unstable tube manifolds 
${\cal W}^{TU}_{B,GL1}$, ${\cal W}^{TU}_{B,GL2}$ remain unaltered in time when 
regarded in the variables $(\xi_B,q_B,\eta_B,p_B)$}. Then, due to the transformation 
(\ref{uxitrabir}), the manifolds transformed back to the original variables have a 
dependence on the angles $\phi_2$, $\phi_3$, etc., implying a dependence on time 
through $M$ independent frequencies. Physically, the manifolds are subject to small 
oscillations (of order $\max{V_i}$, $i=2,3,...$) with respect to a basic static shape 
which is given by their time-invariant form in the variables $(\xi_B,q_B,\eta_B,p_B)$. 
Hence, the manifolds yield spirals with a pattern exhibiting quasi-periodic oscillations 
around the basic spiral patterns induced by the manifolds of the pure bar model. 

\section{Application in a Milky-Way type model}
We now apply the above theory in the case of a Milky Way type galactic model, assuming 
a different pattern speed for the bar and for the spiral arms. We emphasize that this 
is not intended as a modelling of the real spiral structure in the Milky Way, but only 
as a `proof of concept' of the possibility of manifold spirals to support structures 
with more than one pattern speed. 

\subsection{Potential}
We use a variant of the Galactic potential proposed in \cite{petetal2014}, 
which consists of the following components:

{\it Axisymmetric component:} The axisymmetric component is a superposition 
of a disc + halo components, $V_{ax}(\rho,z)=V_d(\rho,z)+V_h(r)$, where 
$r=(x^2+y^2+z^2)^{1/2}$. 
The disc potential has the Miyamoto-Nagai form (\cite{miynag1975}) 
\begin{equation}\label{potdisc}
V_d=\frac{-GM_d}{\sqrt{\rho^2+(a_d+\sqrt{z^2+b_d^2})^2}}
\end{equation}
where $M_d=8.56 \times 10^{10} M\odot$, $a_d=5.3$~kpc and $b_d=0.25$~kpc. The halo 
potential is a $\gamma$-model (\cite{den1993}) with parameters as in \cite{petetal2014}
\begin{equation}\label{pothalo}
V_h=\frac{-GM_h(r)}{r}-\frac{-GM_{h,0}}{\gamma r_h}
\Big[-\frac{\gamma}{1+(r/r_h)^\gamma}+\ln(1+\frac{r}{r_h})^\gamma\Big]_r^{r_{h,max}}
\end{equation}
where $r_{h,max}=100$~kpc, $\gamma=1.02$, and $M_{h,0}=10.7 \times 10^{10} M\odot$, 
and $M_h(r)$ is the function
\begin{equation}\label{mhr}
M_h(r)=\frac{M_{h,0}(r/r_h)^{\gamma+1}}{1+(r/r_h)^\gamma}~~.
\end{equation}

{\it Bar:} The bar potential is as in \cite{lonmur1992} 
\begin{equation}\label{potbar}
V_b=\frac{GM_b}{2a}\ln(\frac{x-a+T_-}{x+a+T_+})
\end{equation}  
with $T_{\pm} = \sqrt{[(a \pm x )^2+y^2+(b+\sqrt{c^2+z^2})^2]}$,  $M_b=6.25 \times 
10^{10} M\odot$, $a=5.25$~kpc, $b=2.1$~kpc and $c=1.6$~kpc. The values of $a,b$ 
set the bar's scale along the major and minor axes in the disc plane ($x$ and $y$ 
respectively), while $c$ sets the bar's thickness in the z-axis (see 
\cite{ger2002}; \cite{ratetal2007}; \cite{caoetal2013}). These values 
where chosen so as to bring the bar's corotation (for $\Omega_{bar}=45$~km/s/kpc) 
to the value (specified by the $L_{1,2}$ points' distance from the center) 
$R_{L1,2} = 5.4$~kpc. Assuming corotation to be at $1.2 - 1.3$ times the bar's 
length, the latter turns to be about $4$Kpc with the adopted parameters. 

{\it Spiral arms}: we use a variant of the logarithmic spiral arms model adopted 
in \cite{petetal2014}. The spiral potential reads (\cite{coxgom2002}) 
\begin{eqnarray}\label{potsp}
\nonumber V_{sp}&=&-4 \pi
{G h_z d_0 C\over K D} \left[sech\left({K z\over\beta}\right)\right]^\beta
F(\rho) \exp\left(-\frac{\rho-\rho_0}{R_s}\right)\nonumber\\  
&\times&\cos (N[\phi-(\Omega_{sp}-\Omega_{bar})t
-\frac{ln(\rho/\rho_0)}{\tan(\alpha)}]) 
\end{eqnarray}
where $N$ is the number of spiral arms and
\begin{equation}
K=\frac{N}{\rho |\sin(\alpha)|}
\end{equation}
\begin{equation}
D=\frac{1+K h_z+0.3(K h_z)^2}{1+0.3K h_z}~,
\end{equation}
\begin{equation}
\beta=K h_z(1+0.4K h_z)~~.
\end{equation}
The function $F(\rho)$ plays the role of a smooth envelope determining the radius  
beyond which the spiral arms are important. We adopt the form 
$F(\rho)= b-c\arctan((R_{s0}-\rho)/kpc)$, with $R_{s0}=6$~kpc, $b=0.474$, $c=0.335$. 
The values of the remaining constants are : $N=2$, $\alpha=-13^\circ$, 
$h_z=0.18$~kpc, $R_s=3$~kpc, $\rho_0=8$~kpc, $C=8/3\pi$. The spiral amplitude is 
determined by setting the value of the density $d_0$. We consider three values, 
namely $d_0=A_{s0}\times 10^8 M_\odot/kpc^3$, with $A_{s0}=1.5$, $3$ and $4$, 
called the weak, intermediate, and strong spirals respectively. These values were 
chosen so as to yield spiral Q-strength values consistent with those reported 
in literature for a mild bar (see \cite{butetal2009}). Our basic model is the 
intermediate one, but as shown below there are only small variations to the basic 
manifold morphology in any of these three choices, since the manifolds' shape is 
determined mostly by the bar. Finally, for the spiral pattern speed we adopt the 
value $\Omega_{spiral} = 20$~km/sec/kpc, which is different from the bar pattern 
speed $\Omega_{bar} = 45$~km/sec/kpc (\cite{ger2011}; \cite{blager2016}).

\begin{figure}
\centering
\includegraphics[scale=0.8]{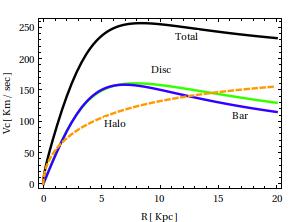}
\caption{The rotation curve (black) corresponding to the potential $V=V_h+V_d+<V_b>$, 
where $<V_b>$ is the $m=0$ (average with respect to all azimuths) part of the bar's 
potential $V_b$. The contribution of each component is shown with different color.} 
\label{fig:rot}
\end{figure}
\begin{figure}
\centering
\includegraphics[scale=0.28]{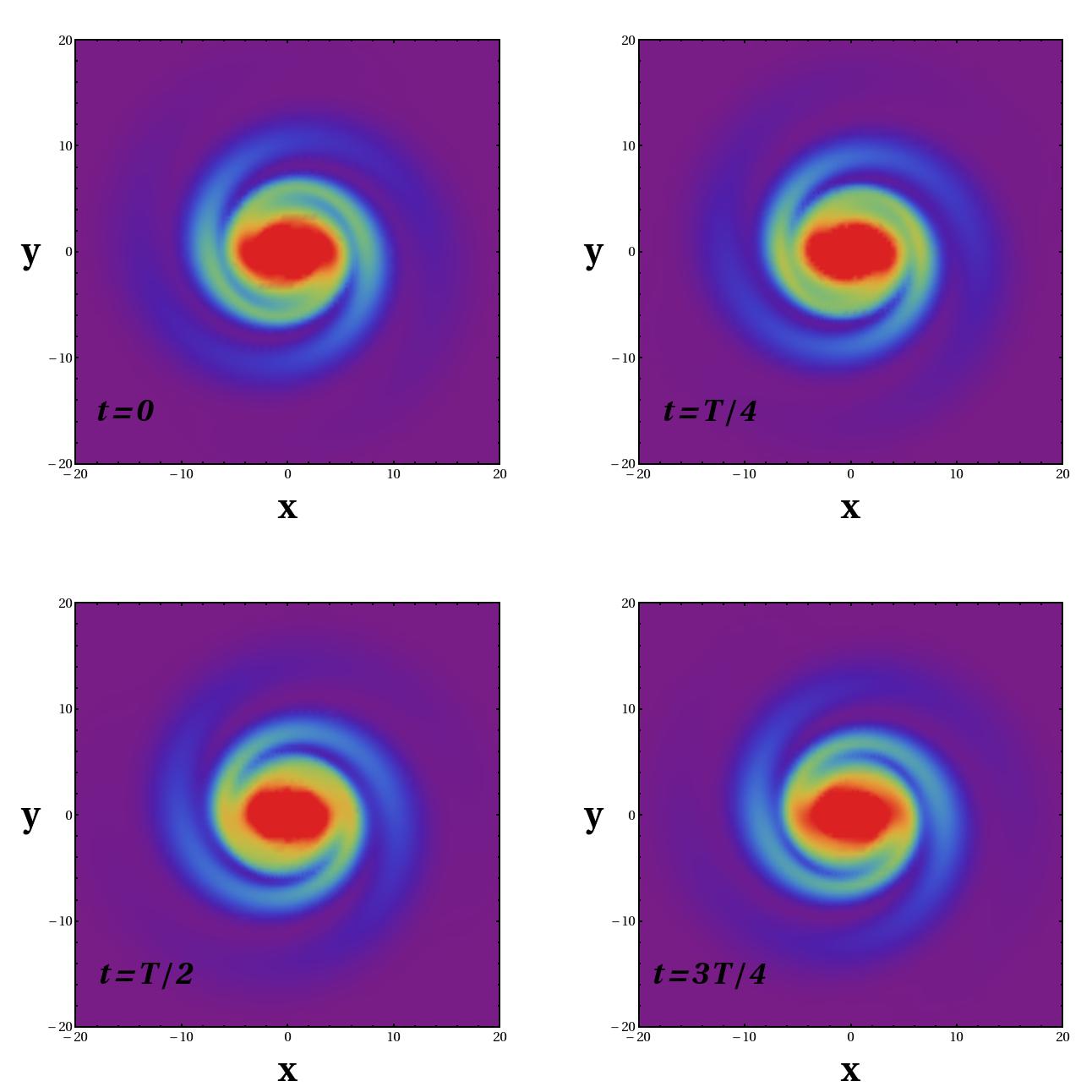}
\caption{A color map of the surface density $\Sigma(\rho,\phi)$ corresponding to the 
potential $V = V_d+V_b+V_{sp}$ (see text), as viewed in the bar's rotating frame, 
at four different snapshots, namely $t=0$ (top left), $t=T/4$ (top right), $t=T/2$ 
(bottom left) and $t=3T/4$ (bottom right), where $T=\pi/|\Omega_{sp}-\Omega_{bar}|$. 
Since $\Omega_{sp}<\Omega_{bar}$, the spirals have a relative clockwise angular 
displacement in time with respect to the bar. However, the morphological continuity 
between bar and spirals is retained in all these snapshots.} 
\label{fig:isoden}
\end{figure}
\begin{figure}
\centering
\includegraphics[scale=1.0]{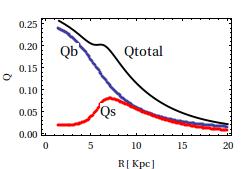}
\caption{The bar, spiral, and total Q-strengths ($Q_b$, $Q_s$ and $Q_{total}$ 
respectively) as functions of the radius $\rho$ in the model including the potential 
terms $V_d$ ,$V_b$ and $V_{sp}$ for the `intermediate' spiral model (see text).} 
\label{fig:qstrength}
\end{figure}
\begin{figure*}
\centering
\includegraphics[scale=0.25]{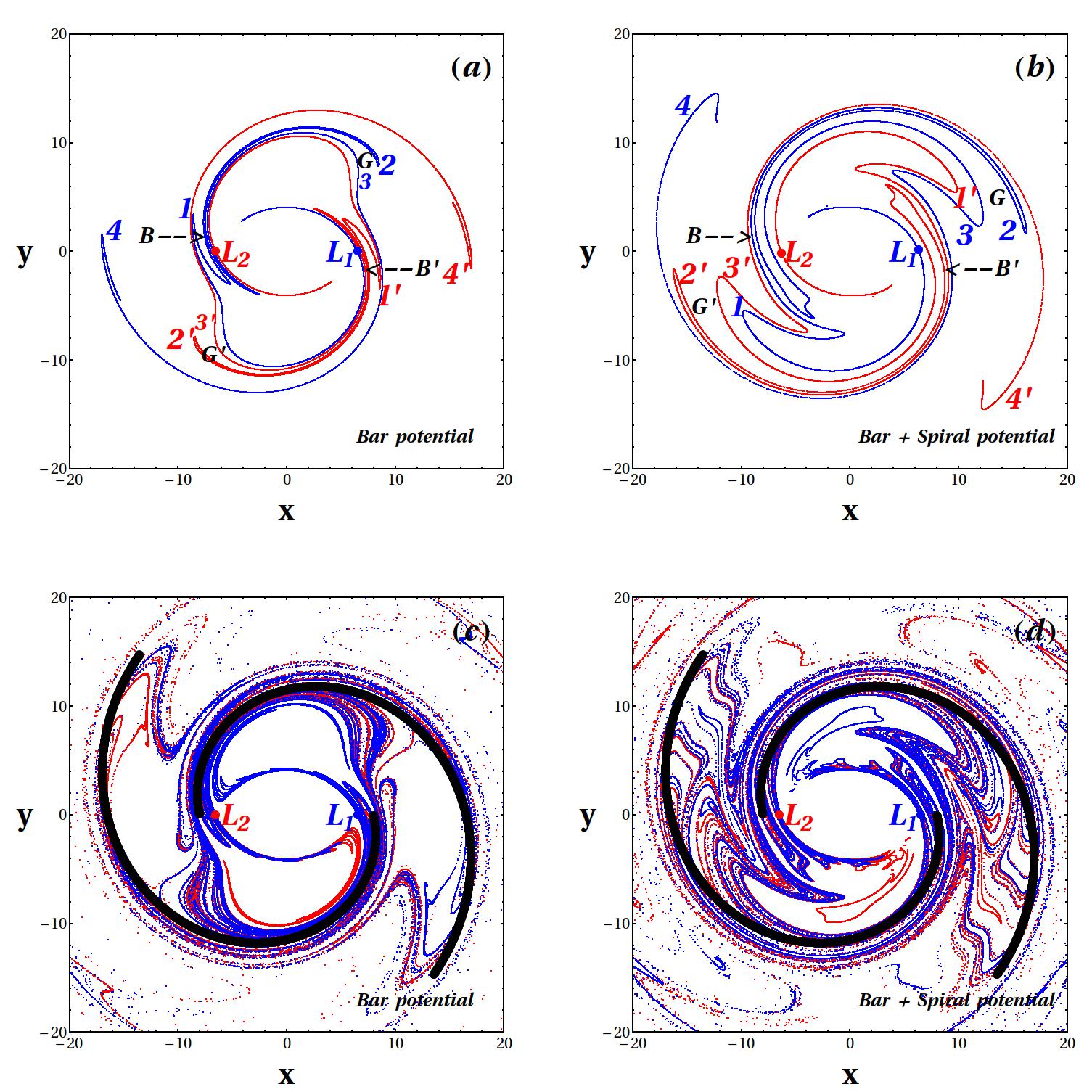}
\caption{(a) The apocentric invariant manifolds in the pure bar model with one pattern 
speed. The manifolds emanating from the points $L_1$ and $L_2$ are plotted with blue 
and red color respectively. (b) The apocentric manifolds if we add the spiral 
potential (intermediate case), assuming that the spirals rotate with the same pattern 
speed as the bar. The addition of the spiral term enhances the structures described 
as `lobes', 'bridges' and 'gaps' (see text). (c) and (d): same as in (a) and (b), 
but with the manifolds computed over a larger length. The black spiral curves 
correspond to the minima of the imposed spiral potential, given by Eq.(\ref{potsp}).} 
\label{fig:mansone}
\end{figure*}

Figure \ref{fig:rot} shows the rotation curve arising from the axisymmetric components 
as well as the azimuthally averaged part of the bar's potential (the corresponding 
component is equal to zero for the spirals).  The model is close to `maximum disc', 
i.e., the rotation curve up to $\sim 10$~kpc is produced essentially by the bar's and 
disc components alone. On the other hand, Figure \ref{fig:isoden} shows an isodensity 
color map of the projected surface density $\sigma(x,y) = \int_{-\infty}^{\infty}
\rho(x,y,z)dz$ in the disc plane, where the density $\rho$ is computed from Poisson's 
equation $\nabla^2V = 4\pi G\rho$ for the potential $V = V_d+V_{bar} + V_{sp}$. The fact 
that the spiral potential has a non-zero relative pattern speed in the bar's frame results 
in a time-dependent spiral pattern in the disc plane. However,  it is well known 
(\cite{selspa1988}) that, under resonable assumptions for the bar and spiral parameters, 
such a time dependence results in a morphological continuity, at most time snapshots, 
between the end of the bar and the spiral arms. For numerically testing the manifold 
theory, we choose below four snapshots as characteristic, corresponding to the times 
$t=0$, $T/4$, $T/2$ and $3T/4$ in Eq.(\ref{potsp}), where $T = \pi/|\Omega_{sp}-
\Omega_{bar}|$. Note that, since the imposed spiral potential has only $\cos2\phi$ 
and $\sin 2\phi$ terms, the spiral patterns shown in Fig.\ref{fig:isoden}, are repeated 
periodically with period $T$. Defining the `phase' of the spirals at a radial distance 
$\rho$ as the angle $\phi_s(\rho)$ where the spiral potential is minimum, given by 
\begin{equation}\label{spphase}
\phi_s(\rho,t) = (\Omega_{sp}-\Omega_{bar})t + (1/\tan(a))\ln(\rho/\rho_0)
~~~ (mod2\pi),
\end{equation} 
we characterize below the relative position of the spirals with respect to the bar by 
the angle $\phi_s(\rho_0,t)$, which is a periodic function of time. The angle $\phi_s$ 
is equal to the angle between the point $L_1$ ($L_2$), which, assuming the bar horizontal, 
lies in the semi-plane $x>0$ ($x<0$), and one of the two local minima of the spiral 
potential at $\rho=\rho_0$, which lies in the semi-plane $y\leq 0$ ($y\geq 0$). 

Regarding the relative bar and spiral contributions to the non-axisymmetric forces, 
figure \ref{fig:qstrength} allows to estimate the relative importance of the 
bar's and spirals' non-axisymmetric force perturbation by showing the corresponding 
Q-strengths as functions of the radial distance $\rho$ in the disc. The Q-strength at 
fixed $\rho$ (e.g. \cite{butetal2009}) is defined for the bar as
\begin{equation}\label{qstr}
Q_b(r)={F_{b,t}^{max}(\rho)\over <F_r(\rho)>} 
\end{equation}
where $F_{b,t}^{max}$ is the maximum, with respect to all azimuths $\phi$, tangential  
force generated by the potential term $V_b$ at the distance $\rho$, while $<F_r(\rho)>$  
is the average, with respect to $\phi$, radial force at the same distance generated 
by the potential $V_d+V_b+V_{sp}$. The bar yields a Q-value $Q_b\approx 0.25$ in its 
inner part which falls to $Q_b\approx 0.15$ to $0.10$ in the domain outside the bar 
where the manifolds (and spirals) develop, i.e., $5$~kpc$<\rho< 10$~kpc. The spirals, 
in turn, yield a maximum $Q_s$ around $\rho\approx 7$~kpc, equal to $Q_s\approx 0.08$ 
in the intermediate model, turning to 0.04 or 0.11 in the weak and strong models 
respectively. Thus, the total Q-strength is about 0.15 to 0.2 in the domain of interest. 

\subsection{Manifold spirals}
A useful preliminary computation regards the form of the apocentric manifolds in the 
above models in two particular cases: i) a pure bar case, and ii) a bar+spiral case, 
assuming, however, the spirals to rotate with the same pattern speed as the bar. The 
corresponding results are shown in Fig.\ref{fig:mansone}. It is noteworthy that even 
the pure bar model yields manifolds which support a spiral response 
(Figs.\ref{fig:mansone}a,c). In addition, the manifolds induce a $R_1$-type ring-like 
structure reminiscent of `pseudorings' (see \cite{but2013} for a review), i.e. rings 
with diameter comparable to the bar's length and a spiral-like deformation with respect 
to a symmetric shape on each side of the bar's minor axis. Adding, now the spiral term 
(with the same pattern speed as the bar) enhances considerably these structures 
(Figs.\ref{fig:mansone}b,d). The most important effect is on the pseudo-ring structure, 
which is now deformed to support the imposed spirals over a large extent. It is of 
interest to follow in detail how the intricate oscillations of the manifolds result 
in supporting the imposed spiral structure. Figure \ref{fig:mansone}b gives the 
corresponding details. We note that the manifolds emanating from the point $L_1$ 
(blue), initially expand outwards, yielding spirals with a nearly constant pitch 
angle. However, after half a turn, the manifolds turn inwards, moving towards the 
neighborhood of the point $L_2$. While approaching there, the manifolds develop 
oscillations, known in dynamics as the `homoclinic oscillations' (see \cite{con2002}) 
for a review). As a result, the manifolds form thin lobes. In Figs.\ref{fig:mansone}a,b 
we mark with numbers 1 to 4 the tips of the first four lobes, and label these lobes 
accordingly. Focusing on Fig.\ref{fig:mansone}b, we note that the lobe 1 is in the 
transient domain between the spirals and the Lagrangian points. However, the lobe 2 
of the manifold emanating from $L_1$ supports the spiral arm originating from the end 
of the bar at $L_2$, and, conversely, the lobe 2' of the manifold emanating from $L_2$ 
supports the spiral arm originating from $L_1$. We call this phenomenon a `bridge' 
(see also \cite{eftetal2019}) and mark the corresponding parts of the manifolds with 
B and B'. One can check that this phenomenon is repeated for higher order lobes of the 
manifolds. Thus, in Fig.\ref{fig:mansone}b, the lobe 3 supports the outer part of the 
pseuroding assosiated with the spiral originating from $L_2$, while lobe 3' 
supports in the same way the spiral originating from $L_1$. Furthermore, between 
lobes 2 and 3 a `gap' is formed (marked G), which separates the pseudoring from 
the outer spiral (and similarly for the gap G' formed between lobes 2' and 3'). 
On the other hand, lobe 4 returns to support the spiral originating from $L_1$. 
Higher order lobes repeat the same phenomenon, but their succession becomes more 
and more difficult to follow, as shown in Fig.\ref{fig:mansone}d. One can remark that 
the manifolds support the spiral geometry mostly in the outer parts of the pseudorings. 
In fact, in the pure bar model we have again the appearance of manifold oscillations, 
leading to lobes, bridges and gaps (Figs.\ref{fig:mansone}a,c), but now the ring part 
is only mildly deformed and clearly separated from the outer lobes which support 
spirals. 
 
\begin{figure*}
\centering
\includegraphics[scale=0.25]{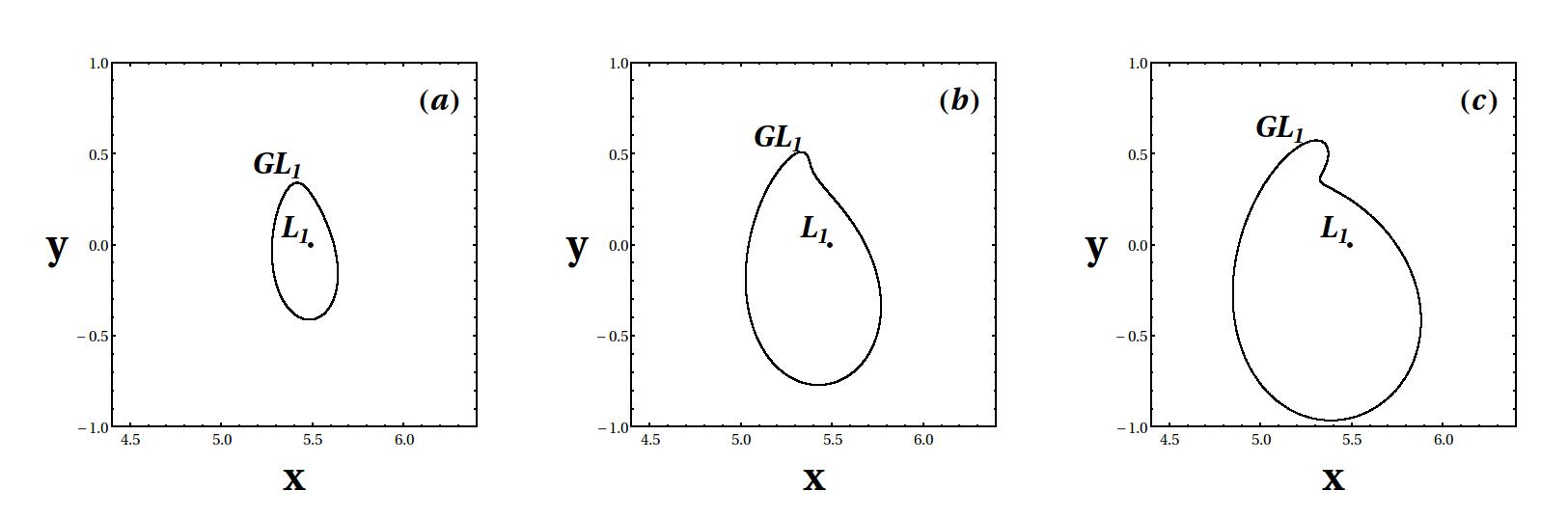}
\caption{The periodic orbit $GL_1$ in the weak (left), intermediate (center) 
and strong (right) spiral cases. The size of the orbit increases with the spiral 
amplitude. } 
\label{fig:perorb}
\end{figure*}

We now examine how these morphologies are altered, if, instead, we assume the spirals 
to rotate with a different pattern speed than the bar. The computation of the manifolds 
in this case proceeds along the steps described in section 2. For the computation of 
the initial diagonalizing transformation matrix ${\cal A}$ (Eq.\ref{lintra}) as well 
as the canonical transformation (\ref{uxitra2}) we proceed as described in the Appendix. 
In particular, we use the Lie series method in order to perform all series computations.  
These series allow us to compute initial conditions for the periodic orbits $GL_1$ and 
$GL_2$ (Eq.(\ref{uxitral1})).  Finally, we numerically refine the latter computation 
using Newton-Raphson to obtain the periodic orbits with many significant figures. 
More specifically, since the potential depends periodically on time 
(with period $T=\pi/(\Omega_{bar}-\Omega_{sp})$), we consider a stroboscopic map 
\begin{equation}\label{strob}
(x(0),y(0),p_\rho(0),p_\phi(0))\rightarrow(x(T),y(T),p_\rho(T),p_\phi(T))
\end{equation}
which maps any initial condition at the time $t=0$ to its image at the time $t=T$ under 
the full numerical equations of motion without any approximation. Then, the periodic 
orbits $GL_1$ and $GL_2$ are fixed points of the above map. As shown in 
Fig.\ref{fig:perorb}, the periodic orbits $GL_1$ and $GL_2$ found by the above method 
form epicycles around the Lagrangian points $L_1$ and $L_2$ of the pure bar model. 
However, the orbits $GL_1$ and $GL_2$ should not be confused with the epicyclic 
Lyapunov orbits $PL_1$, $PL_2$ used in past manifold calculations in models with one 
pattern speed (\cite{vogetal2006}). In particular, the orbits $PL_1$ and $PL_2$ exist 
as a family of orbits in a fixed bar model, whose size depends continuously on the value 
of the Jacobi energy $E_J>E_{L1}$. Under specific conditions, the orbits $PL_{1,2}$ can 
be generalized to 2D-tori in the case of one extra pattern speed. However, this 
generalization requires the use of Kolmogorov-Arnold-Moser theory (\cite{kol1954}; 
\cite{arn1963}; \cite{mos1962}) which is beyond our present scope. On the contrary, 
in the two-pattern speed case, for a fixed choice of the potential $V_2=V_{sp}$ 
(Eq.\ref{potsp}) and $\Omega_2=\Omega_{sp}$ there exist {\it unique} $GL_1$ and $GL_2$ 
orbits, which generalize the unique Lagrangian points of the corresponding pure bar 
model. In fact, the orbits of Fig.\ref{fig:perorb} have relative size of the order 
of the ratio of the $m=2$ Fourier amplitudes of the bar and of the spiral potential 
at the radius $\rho=\rho_{L1,2}$. This is about $0.5$~kpc, $1.2$~kpc and $1.5$~kpc 
in the weak, intermediate and strong spiral case respectively. 
\begin{figure*}
\centering
\includegraphics[scale=0.32]{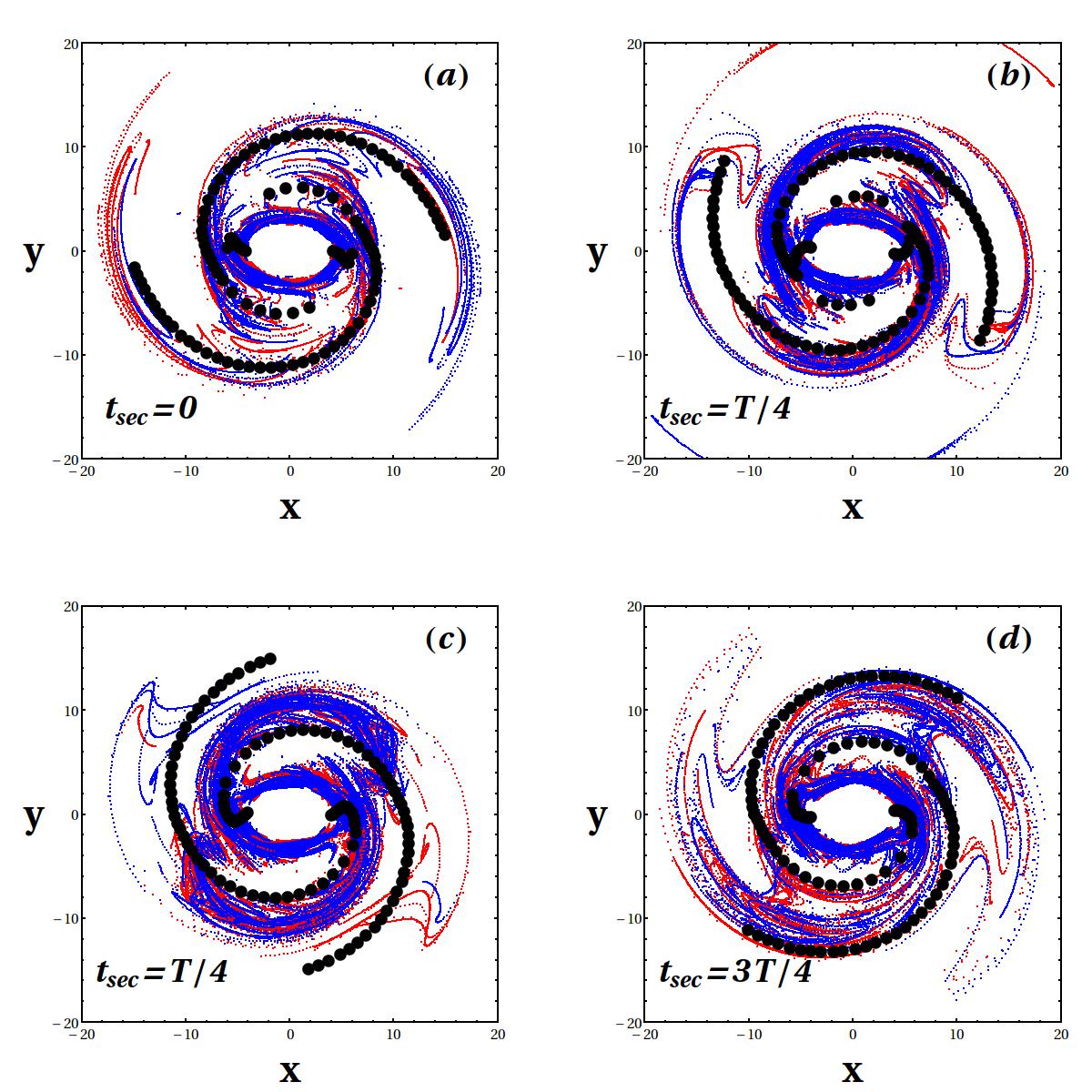}
\caption{The `double section' apocentric manifolds ${\cal W}^{AU}_{GL1}(t_{sec})$ 
(blue) and ${\cal W}^{AU}_{GL2}(t_{sec})$ (red) are plotted at four different times 
$t_{sec}$ as indicated in each panel. The black-dotted curves mark the local maxima 
of the surface density $\sigma(\rho,\phi,t)$ corresponding to the potential 
$V=V_d(\rho)+V_b(\rho,\phi)+V_{sp}(\rho,\phi,t)$ at the times $t=t_{sec}$. These 
maxima are plotted in the domain $6$~kpc$<\rho<15$~Kpc, where the imposed spirals 
have a significant amplitude.} 
\label{fig:apotwopat30}
\end{figure*}
\begin{figure*}
\centering
\includegraphics[scale=0.35]{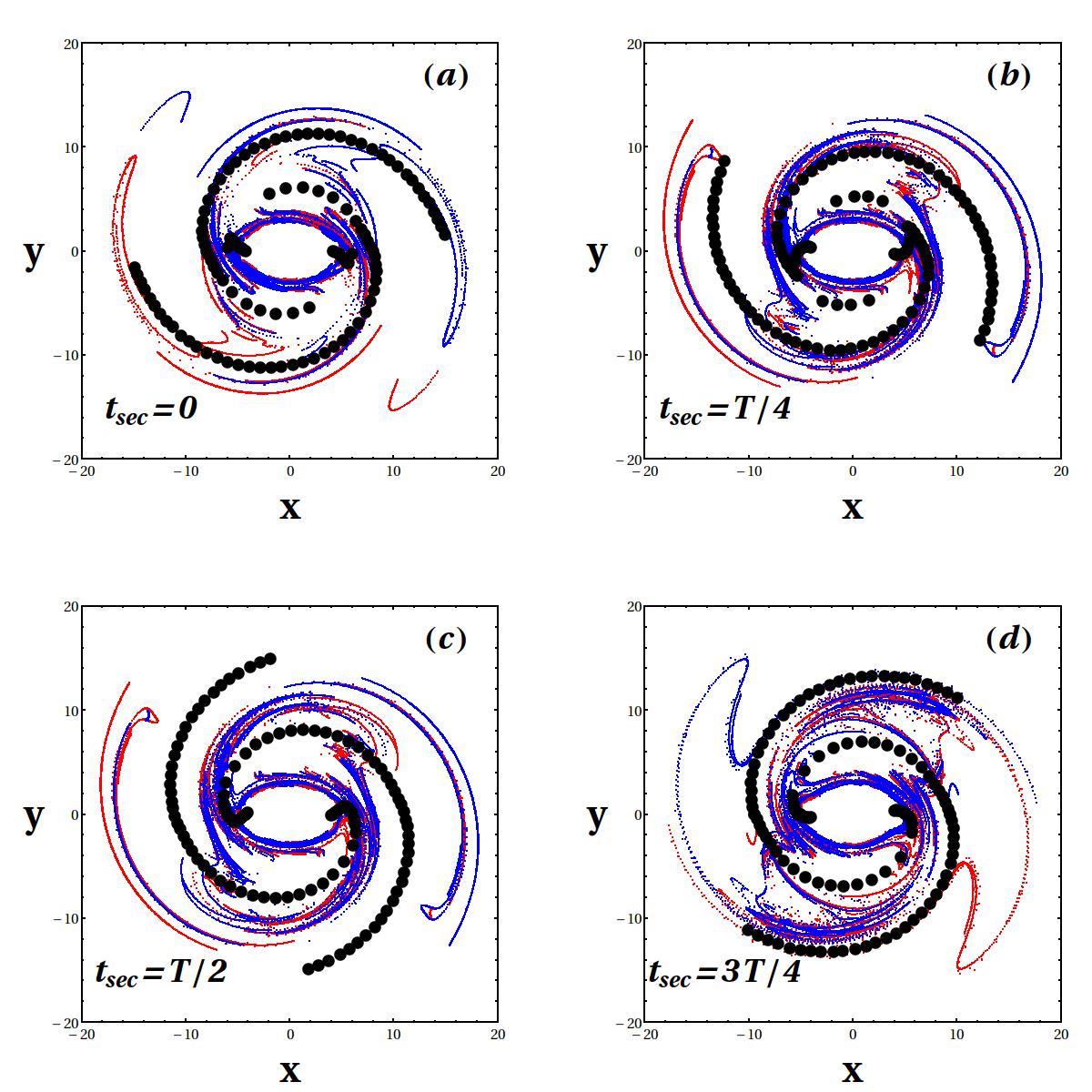}
\caption{Same as in Fig.\ref{fig:apotwopat30}, but for the weak spiral model.} 
\label{fig:apotwopat15}
\end{figure*}
\begin{figure*}
\centering
\includegraphics[scale=0.35]{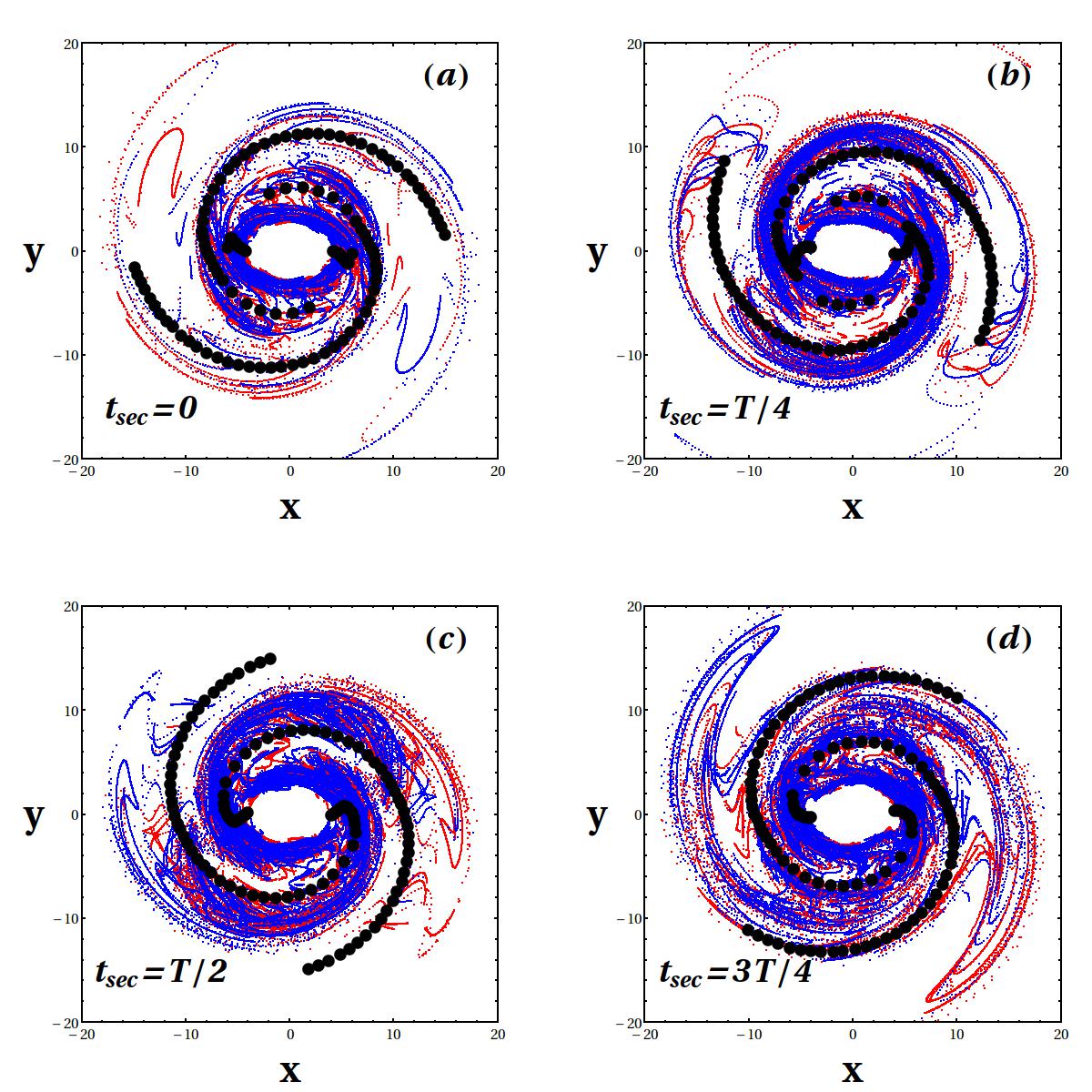}
\caption{Same as in Fig.\ref{fig:apotwopat30}, but for the strong spiral model.} 
\label{fig:apotwopat40}
\end{figure*}

The computation of the unstable manifolds of the orbits $GL_1$ and $GL_2$ is now 
straightforward: focusing on, say, $GL_1$, we first compute the $4\times 4$ variational 
matrix $\Lambda$ of the mapping (\ref{strob}) evaluated at the fixed point of the 
periodic orbit $GL_1$. The matrix $\Lambda$ satisfies the symplecticity condition 
$\Lambda\cdot{\cal J}\cdot\Lambda^T = \Lambda^T\cdot {\cal J}\cdot\Lambda = 
{\cal J}$, and it has two real reciprocal eigenvalues $\lambda_1,\lambda_2=
1/\lambda_1$, with $|\lambda_1|>1$, and two complex congugate ones with unitary 
measure $\lambda_{3,4}= e^{\pm i\omega T}$ for some positive $\omega$. Denoting 
by $e^U_{GL1}$ the unitary eigenvector of $\Lambda$ associated with the eigenvalue 
$\lambda_1$, we then consider a small segment divided in $10^5$ initial conditions 
of the form $(\rho_{i,0},\phi_{i,0},p_{\rho,i,0},p_{\phi,i,0}$, $i=1,...,100000$ 
defined by $(\rho_{i,0},\phi_{i,0},p_{\rho,i,0},p_{\phi,i,0}) = 
(\rho_{GL1}+\delta\rho_{i,0},\phi_{GL1}+\delta\phi_{i,0},
p_{\rho,GL1}+\delta p_{\rho,i,0},p_{\phi,GL1}+\delta p_{\phi,i,0})$ where 
$(\delta\rho_{i,0},\delta\phi_{i,0},\delta p_{\rho,i,0},\delta p_{\phi,i,0}) = 
(i/100000)\times \Delta S\times e^U_{GL1}$, with $\Delta S=0.001$. Propagating 
all these orbits forward in time yields an approximation of the unstable flux-tube 
manifold ${\cal W}^{TU}_{GL1}$ (see section 2). 

In contrast to what happens in the one-pattern speed model, under the presence 
of the second pattern speed the projection of the `flux-tube' manifolds 
${\cal W}^{TU}_{GL1}$ in the disc plane varies in time. In order to efficiently 
visualize how the manifolds develop in space and time, in the following plots we 
use an {\it apocentric double section} of the manifolds, denoted 
${\cal W}^{AU}_{GL1}(t_{sec})$, which depends on a chosen value of the `section 
time' $t_{sec}$. The apocentric double section for a given time $t_{sec}$ is defined 
as follows: keeping track of all the points of the tube manifolds generated by the above 
initial conditions, we retain those points corresponding to integration times 
$t=nT +t_{sec}\pm\Delta T$, with $n=0,1,2,\ldots$ and $\Delta T$ small ($\Delta T = 0.1T$ 
in all our calculations), and at the same time satisfying the apocentric condition 
$\dot{p}_\rho\simeq 0$ (with an accuracy defined by $|\dot{\rho}|<|\ddot{\rho}|dt$ 
where $|\ddot{\rho}|$ is the measure of the radial acceleration at the 
evaluation point, and $dt=0.001T$ is the integration timestep).  This representation 
allows to obtain the intersections of the manifolds with an apocentric surface of 
section (see \cite{eft2010} for a discusion of how the apocentric manifolds compare 
with the full flux-tube manifolds). However, it also allows to capture the dependence 
of the form of the manifolds on {\it time}, through the chosen value of $t_{sec}$. 

Figure \ref{fig:apotwopat30} shows the main result: the manifolds 
${\cal W}^{AU}_{GL1}(t_{sec})$ (blue points) and ${\cal W}^{AU}_{GL2}(t_{sec})$ (red 
points) computed as above, are shown at four different times $t_{sec}$, namely 
$t_{sec}=0$, $T/4$, $T/2$ and $3T/4$, corresponding to the same snapshots as in 
Fig.(\ref{fig:isoden}). The spiral phase $\phi_s(\rho_0,t_{sec})$ has the values 
$0$, $-\pi/4$, $-\pi/2$ and $-3\pi/4$, respectively. The black dotted curves 
superposed to the manifolds correspond to the maxima of the surface density in the 
annulus $6~kpc<\rho< 15~kpc$, as found from the data of Fig.\ref{fig:isoden}.  
These figures repeat periodically after the time $t_{sec}=T$. 

The key result from Fig.\ref{fig:apotwopat30} is now evident: The spiral maxima rotate 
clockwise with respect to the bar (with angular velocity equal to $2\pi/(\Omega_{bar}
-\Omega_{sp}$). The manifolds ${\cal W}^{AU}_{GL1,2}$ adapt their form to the rotation  
of the spiral maxima, thus acquiring a time-varying morphology. In particular, 
the manifolds always form bridges and gaps, thus supporting a pseudo-ring as well as 
an outer spiral pattern. The spiral-like deformation of the pseudo-ring 
is most conspicuous at $t_{sec}=0$, corresponding to a spiral phase $\phi_s(\rho_0)=0$, 
and it remains large at the times $t_{sec}=T/4$ and $3T/4$, i.e., at the spiral phases 
$\phi_s(\rho_0)=-\pi/4$ and $-3\pi/4$. At all these phases the spiral maxima at 
$\rho=\rho_0$ remain close to the bar's major axis, thus the manifolds tend to take 
a form similar to the one of Fig.\ref{fig:mansone}d (in which $\phi_s(\rho_0)=0$ 
always since we set $\Omega_{bar}=\Omega_{sp}$). On the other hand, at $t_{sec}=T/2$, 
($\phi_s(\rho_0)=-\pi/2$) the spiral maxima at $\rho=\rho_0$ are displaced by an 
angle $\pi/2$ with respect to the bar's horizontal axis. Then, the manifolds yield 
more closed pseudo-rings, and they temporarily stop supporting the imposed spirals. 
Comparing the three phases $\phi_s(\rho_0)=0$, $-\pi/4$ and $-3\pi/4$ we find that 
the agreement between manifolds and imposed spirals is best at the phases 
$\phi_s(\rho_0)=0$ and $-3\pi/4$, while the manifolds support the imposed spiral 
mostly in their pseudo-ring part at $\phi_s(\rho_0)=-\pi/4$. 

Altering the spirals' amplitude (Figs.\ref{fig:apotwopat15} and \ref{fig:apotwopat40}) 
makes no appreciable difference to the above picture. The main noticed difference 
regards the thickness of the manifolds' lobes, which increases with the imposed 
spiral amplitude, since, in general, the manifolds make larger oscillations near 
the bridges when the non-axisymmetric perturbation increases. This means also that 
the trajectories supporting these spirals are more chaotic. 

\subsection{Discussion}
Commenting on the loss of support of the manifolds to the imposed spiral maxima near 
$\phi_s(\rho_0)=-\pi/2$, we remark that under the scenario that the manifolds provide 
the backbone supporting chaotic spirals, a temporary loss of support implies that the 
spiral response to the manifolds should have its minimum strength when the spirals have 
a relative phase $\pm\pi/2$ with respect to the bar's major axis. Since the bar-spiral 
relative configuration (and the manifolds' shape) is repeated periodically, with 
period $T=\pi/(\Omega_{bar}-\Omega_{sp})$, we conclude that under the manifold scenario 
the amplitude in the response spiral should exhibit periodic time variations, with a 
period equal to $T$, i.e., the manifolds support {\it recurrent spirals} with 
the above periodicity. The appearance of recurrent spirals in multi-pattern speed 
N-body models is well known (see \cite{selwil1993}, \cite{sel2003}). The manifold 
theory provides a specific prediction about the period of the recurrence, which is 
testable in such experiments by the time-Fourier analysis of the non-axisymmetric 
patterns.  On the other hand, the picture presented above is still `static', in the 
sense that it does not take into account phenomena which alter in time the imposed 
non-axisymmetric modes. Such phenomena are nonlinear interactions between distinct 
modes, and the enhancement or decay of the spirals associated with disc instabilities 
(e.g. swing amplification) or with dissipation mechanisms (e.g. disc heating at 
resonances or gas phenomena). In all such circumstances, the manifolds provide a 
way to understand the behavior of chaotic trajectories beyond the bar. Thus, a full 
exploration of the connection between manifolds and collective disc phenomena is 
proposed for further study.

\section{Conclusions}
In the present study we have examined the possibility that manifold spirals in barred 
galaxies are consistent with the presence of multiple pattern speeds in the galactic 
disc. In section 2 we gave the main theory, and in section 3 we constructed numerical 
examples of such manifold spirals. Our main conclusions are the following:

1. In the case of one pattern speed, the basic manifolds are those generated by the 
unstable manifolds of the Lagrangian points $L_1$ and $L_2$. In the case of multiple 
pattern speeds, it can be established theoretically (see section 2) that, while 
Lagrangian equilibrium points no longer exist in the bar's rotating frame, such points 
are replaced by `generalized Lagrangian orbits' (the orbits $GL_1$ and $GL_2$) which 
play a similar role for dynamics. These orbits are periodic, with period equal to 
$\pi/|\Omega_{sp}-\Omega_{bar}|$, if there is one spiral pattern rotating with speed 
$\Omega_{sp}$ different from $\Omega_{bar}$. If there are more than one extra 
patterns, with speeds $\Omega_2$, $\Omega_3$, etc., the generalized orbits $GL_1$ 
and $GL_2$ perform epicycles around the Lagrangian points $L_1$ and $L_2$ of the 
pure bar model with incommensurable, in general, frequencies 
$|\Omega_i-\Omega_{bar}|$, $i=1,2,...$. Furthermore, in all cases the orbits 
$GL_1$ and $GL_2$ are simply unstable, a fact implying that they possess unstable 
manifolds ${\cal W}_{GL1}$, ${\cal W}_{GL2}$. When the extra patterns have a 
small amplitude with respect to the bar's amplitude, perturbation theory establishes 
that the manifolds ${\cal W}_{GL1}$, ${\cal W}_{GL2}$ undergo small time variations 
(with the same frequencies $|\Omega_i-\Omega_{bar}|$, $i=1,2,...$), but their 
form in general exhibits only a small deformation with respect to the manifolds 
${\cal W}_{L1}$, ${\cal W}_{L2}$ of the pure bar model. Thus, the manifolds 
${\cal W}_{GL1}$, ${\cal W}_{GL2}$ support trailing spiral patterns. 

2. In section 3 we explore a simple bar-spiral model for a galactic disc with 
parameters relevant to Milky-Way dynamics. In this model we construct manifold spirals 
in both cases of a unique pattern speed ($\Omega_{sp}=\Omega_{bar}$) or two distinct 
pattern speeds ($\Omega_{sp}<\Omega_{bar}$). The pure bar model already generates  
manifolds which support spiral patterns, and also an inner ring around the bar. 
Imposing further spiral perturbations in the potential generates mostly a deformation 
of the manifolds, with the ring evolving to a spiral-like `pseudo-ring'. The spiral and 
ring structures generated by the manifolds connect to each other through `bridges' 
(see Figs.\ref{fig:mansone},\ref{fig:apotwopat30}, \ref{fig:apotwopat15}, and 
\ref{fig:apotwopat40}). This implies that, after a `bridge', the manifold emanating 
from the neighborhood of the bar's $L_1$ ($L_2$) point supports the spiral arm 
associated with the bar's end near the $L_2$ ($L_1$) point. From the point of view 
of dynamics, these connections are a manifestation of {\it homoclinic chaos}. 

3. We find that the manifold theory gives good fit to at least some part of to the 
imposed spirals in both the single and multiple pattern speed models. Focusing on 
numerical examples in which the spiral and bar pattern speeds satisfy 
$\Omega_{sp}<\Omega_{bar}$, the main behavior of the manifold spirals can be 
characterized in terms of the (time-varying) phase $\phi_s(\rho_0)$ (Eq.(\ref{spphase})). 
The manifolds support the imposed spirals over all the latter's length at phases 
$\phi_s(\rho_0)=0$ or $-3\pi/4$, and they support mostly pseudo-ring like spirals 
near the phase $-\pi/4$ (the phase $\phi_s$ is negative since the spirals have a 
retrograde relative rotation with respect to the bar). On the other hand, the manifolds 
deviate from the imposed spirals near the phase $-\pi/2$. Both the manifolds' shape 
and the imposed bar-spiral relative configuration are repeated periodically, with 
period $T=\pi/(\Omega_{bar}-\Omega_{sp})$. Thus, we argued that the temporary loss 
of support of the manifolds to the imposed spirals suggests a natural period for 
recurrent spirals, equal to $T$. 
 
In summary, our analysis shows that {\it manifold spirals in galactic discs are, in 
general, consistent, with the presence of multiple pattern speeds}. Nevertheless, 
the manifold spirals in this case oscillate in time, thus, they support the imposed 
spirals along a varying length, which fluctuates from small to almost complete, 
depending on the relative phase of the spirals with respect to the bar. The manifolds 
also produce ring and pseudo-ring structures, morphologically connected to the 
spirals via the phenomenon of `bridges' (section 3). These features are present 
in real galaxies (\cite{but2013}), but testing their connection to manifolds in 
specific cases of galaxies requires a particular study. \\
\\
{\bf Acknowledgements:} We acknowledge support by the Research Committee of the 
Academy of Athens through the grant 200/895. C.E. acknowledges useful discussions 
with Dr. E. Athanassoula.

\appendix
\section{Series construction}
Starting from the Hamiltonian (\ref{hamext}), we implement the method of  
composition of Lie series in order to arrive to the Hamiltonian (\ref{hamextxieta}), 
by the following steps:\\
\\
1) {\it Expansion}: We compute the Lagrangian points $L_1$, $L_2$ of the Hamiltonian 
(\ref{hambar}).  Selecting, say, the point $L_1$, with co-ordinates 
$\rho_{L1},\phi_{L_1},0,p_{\phi,L1}$, we expand the full Hamiltonian (\ref{hamext}) 
in a polynomial series in the variables $\delta\rho = \rho-\rho_{L1}$, 
$\delta\phi=\phi-\phi_{L1}$, $J_\phi=p_\phi-p_{\phi,L1}$ (see section 2). In a 
computer-algebraic implementation, all expansions are carried up to a maximum 
truncation order $N_t$, set as $N_t=10$.\\
\\
2) {\it Diagonalization:} From the quadratic part $H_{0,2}$ of the Hamiltonian we 
compute the variational matrix ${\cal M}$ at $L_1$, as in Eq.(\ref{varmat}), as well 
as its eigenvalues $\lambda_{1,3}=\pm\nu$, $\lambda_{2,4}=\pm i\kappa$, with 
$\nu,\kappa>0$, and associated eigenvectors $e_i$, $i=1,\ldots,4$. Each eigenvector 
has four components, thus it can be written as a $4\times 1$ column vector. We then 
form the $4\times 4$ matrix ${\cal B}=(c_1e_1,c_2e_2,c_1e_3,c_2e_2)$, with unspecified 
coefficients $c_1,c_2$, which contains the four vectors as its columns (multiplied 
by the $c_i$'s). Applying the symplectic condition ${\cal B}^T \cdot {\cal J} 
\cdot {\cal B} = {\cal J}$, where ${\cal J}_4$ is the $4\times 4$ fundamental
symplectic matrix
\begin{equation}
{\cal J}_4=
\begin{bmatrix}
    0       & I_2 \\
    -I_2    & 0 \\
\end{bmatrix}
\end{equation}
with $I_2$ equal to the $2\times 2$ identity matrix, yields two independent equations 
allowing to specify the coefficients $c_1,c_2$, and hence all the entries of the 
constant matrix ${\cal B}$. This matrix ${\cal A}$ in the transformation (\ref{lintra}) 
is then given by ${\cal A} = {\cal B}\cdot{\cal C}$ where
\begin{equation}
{\cal C}=
\begin{bmatrix}
1                  & 0                & 0               & 0                \\
0                  &{1\over\sqrt{2}}  & 0               &{-i\over\sqrt{2}} \\
0                  & 0                & 1               & 0                \\
&-{i\over\sqrt{2}} & 0                &{1\over\sqrt{2}} & 0                \\
\end{bmatrix}
\end{equation}

3) {\it Normalization using Lie series:} We use the Lie method of normal form 
construction (see \cite{eft2012} section 2.10 for a tutorial) in order to pass  
from the Hamiltonian (\ref{hamextuv}) to the Hamiltonian (\ref{hamextxieta}). 
Briefly, we consider a sequence of canonical transformations 
$(u^{(r-1)},Q^{(r-1)},v^{(r-1)},P^{(r-1)})$ $\rightarrow$ 
$(u^{(r)},Q^{(r)},v^{(r)},P^{(r)})$, with $r=1,2,\ldots,N_t$, where 
$(u^{(0)},Q^{(0)},v^{(0)},P^{(0)})$ $\equiv$ $(u,Q,v,P)$ and 
$(u^{(N_t)},Q^{(N_t)},v^{(N_t)},P^{(N_t)})$, $\equiv$ $(\xi,q,\eta,p)$ 
defined through suitably defined generating functions 
$\chi_1,\chi_2,\ldots,\chi_{N_t}$, through the recursive relations
\begin{eqnarray}\label{cantra}
(u^{(r-1)},Q^{(r-1)},v^{(r-1)},P^{(r-1)}) =~~~~~~~~~~~~~~~ \\ 
~~~~~~~~~~~~~~~~~\exp\left({\cal L}_{\chi_r}\right)
(u^{(r)},Q^{(r)},v^{(r)},P^{(r)}) \nonumber
\end{eqnarray}
where ${\cal L}_{\chi_r}$ denotes the Poisson bracket operator
${\cal L}_{\chi_r}\cdot = \{\cdot,\chi_r\}$, and $\exp({\cal L}_{\chi_r}) = 
\sum_{j=0}^{\infty}{\cal L}_{\chi_r}^j/j!$ (truncated at order $N_t$. Once the 
involved generating functions $\chi_r$ $r=1,2,\ldots,N_t$ are specified, 
Eq.(\ref{cantra}) allows to define the transformation of Eq.(\ref{uxitra2}), and 
hence the periodic orbit $GL_1$ through Eq.(\ref{uxitral1}). 

It remains to see how to compute the functions $\chi_r$. This is accomplished via a 
recursive algorithm, allowing to transform the original Hamiltonian $H^{(0)}\equiv H$, 
with $H$ given by Eq.(\ref{hamextuv}) to its final form $H^{(N_t)}$ given by 
Eq.(\ref{hamextxieta}). We consider the $r-$th normalization step, and give explicit 
formulas in the case of one extra pattern speed, in which we have one extra angle 
$\phi_2=\phi_s$ (generalization to $M$ extra pattern speeds is straightforward). 
The Hamiltonian has the form 
\begin{equation}\label{hamrm1}
H^{(r-1)} = Z_2+Z_3+...+Z_{r+1} + 
R_{r+2}^{(r-1)}+\ldots+R_{N_t+2}^{(r-1)}
\end{equation}
where i) subscripts refer to polynomial order in the variables 
$(u^{(r-1)},Q^{(r-1)},v^{(r-1)},P^{(r-1)})$, and ii) the terms $Z_i$, $i=2,\ldots,r+1$ 
are `in normal form', i.e., contain no monomials linear in $(u^{(r-1)},v^{(r-1)})$. 
The remainder term $R_{r+2}^{(r-1)}$ has the form  
$$
R_{r+2}^{(r-1)} = \sum_m\sum_{k_1+k_2+l_1+l_2=r+2} a_{k_1,k_2,l_1,l_2}^{(r-1)}
$$
$$
\times
e^{i m\phi_s}
(u^{(r-1)})^{k_1},(Q^{(r-1)})^{k_2},(v^{(r-1)})^{l_1},(P^{(r-1)})^{l_2}
$$
with $k_1,k_2,l_1,l_2\geq 0$, and $m$ integer. Then, the generating function $\chi_r$ 
is given by
\begin{eqnarray}\label{chir}
\chi_r &=&  
\sum_{k_1+l_1=1} a_{k_1,k_2,l_1,l_2}^{(r-1)} \\
&\times& 
{e^{i m\phi_s}
(u^{(r)})^{k_1},(Q^{(r)})^{k_2},(v^{(r)})^{l_1},(P^{(r)})^{l_2}
\over
(l_1-k_1)\nu+i[(k_2-l_2)\kappa+m(\Omega_s-\Omega_{bar})]}~~.\nonumber
\end{eqnarray}
With the above rule, the Hamiltonian becomes `in normal form' up to the terms 
of polynomial order $r+2$, namely
\begin{eqnarray}\label{hamrm}
H^{(r)} &=& \exp({\cal L}_{\chi_r})H^{(r-1)} \\
&=& Z_2+Z_3+...+Z_{r+2} + 
R_{r+3}^{(r)}+\ldots+R_{N_t+2}^{(r)}~~.   \nonumber
\end{eqnarray}
Hence, repeating the procedure $N_t$ times leads to the Hamiltonian (\ref{hamextxieta}).

\end{document}